\documentclass[%
reprint,
superscriptaddress,
aps,
pra,
longbibliography
]{revtex4-1}

\usepackage{graphicx} 
\usepackage{dcolumn} 
\usepackage{bm} 
\usepackage{braket}
\usepackage{amsmath}
\usepackage{bm}
\usepackage{amssymb}
\usepackage{setspace}
\usepackage[euler]{textgreek}
\allowdisplaybreaks
\usepackage{float}

\usepackage{multirow}
\usepackage{tabularx}
\usepackage[colorlinks=true,bookmarks=false,citecolor=blue,urlcolor=blue]{hyperref}

\newcommand*\pdistant{$p_{\text{distant}}$}
\newcommand*\plocal{$p_{\text{local}}$}

\begin{document}
\title{A Quantum Router Architecture for High-Fidelity Entanglement Flows in Quantum Networks}

\author{Yuan Lee}
\affiliation{Department of Electrical Engineering and Computer Science, Massachusetts Institute of Technology, Cambridge, Massachusetts 02139, USA}

\author{Eric Bersin}
\affiliation{Department of Electrical Engineering and Computer Science, Massachusetts Institute of Technology, Cambridge, Massachusetts 02139, USA}

\author{Axel Dahlberg}
\affiliation{QuTech, Delft University of Technology, and Kavli Institute of Nanoscience
Delft, The Netherlands}

\author{Stephanie Wehner}
\affiliation{QuTech, Delft University of Technology, and Kavli Institute of Nanoscience
Delft, The Netherlands}

\author{Dirk Englund}
\email{englund@mit.edu}
\affiliation{Department of Electrical Engineering and Computer Science, Massachusetts Institute of Technology, Cambridge, Massachusetts 02139, USA}
\affiliation{Research Laboratory of Electronics, Massachusetts Institute of Technology, Cambridge, Massachusetts 02139, USA}

\begin{abstract}
The past decade has seen tremendous progress in experimentally realizing the building blocks of quantum repeaters. Repeater architectures with multiplexed quantum memories have been proposed to increase entanglement distribution rates, but an open challenge is to maintain entanglement fidelity over long-distance links. Here, we address this with a quantum router architecture comprising many quantum memories connected in a photonic switchboard to broker entanglement flows across quantum networks. We compute the rate and fidelity of entanglement distribution under this architecture using an event-based simulator, finding that the router improves the entanglement fidelity as multiplexing depth increases without a significant drop in the entanglement distribution rate. Specifically, the router permits channel-loss-invariant fidelity, i.e. the same fidelity achievable with lossless links. Furthermore, this scheme automatically prioritizes entanglement flows across the full network without requiring global network information. The proposed architecture uses present-day photonic technology, opening a path to near-term deployable multi-node quantum networks. 
\end{abstract}
\maketitle

\section*{Introduction}
Quantum networks distribute quantum information to enable functions that are impossible on classical networks. Key to these applications is the sharing of entanglement between many users over large distances, allowing quantum key distribution, distributed quantum computing, and quantum-enhanced sensing. While entanglement distribution has been demonstrated over short distances~\cite{Hensen_2015}, long-distance quantum networking is hampered by the exponential loss of photons in optical fibers~\cite{Pirandola_2017}. Quantum repeaters~\cite{Briegel_1998} can overcome this problem by forming chains of entangled nodes.

\begin{figure*}[t]
\centering
\includegraphics[scale = 1]{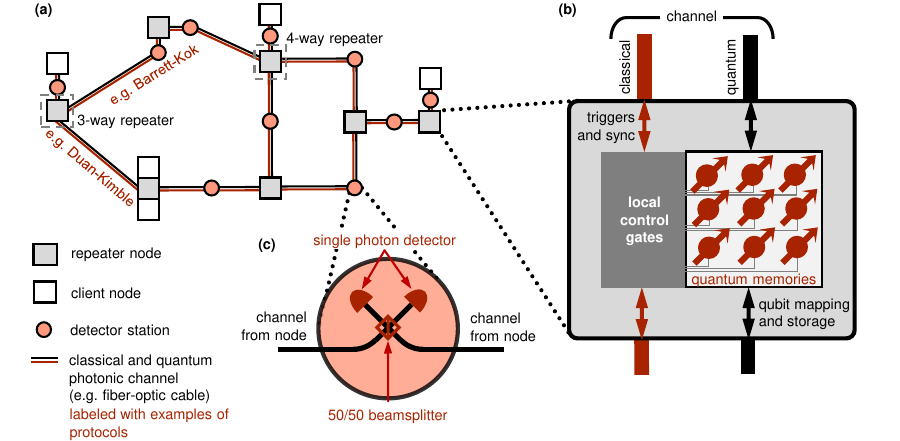}
\caption{Components of a quantum repeater network. The quantum network comprises client nodes, which end users directly access, and repeater nodes, which connect clients by propagating entanglement through the network. \textsf{\textbf{(a)}}~Physical layer implementation. In the physical layer~\cite{Dahlberg_2019}, repeater nodes may be connected by a variety of entanglement protocols. While all protocols require transmission of an entangled photon though a photonic channel, the specific protocol used in each link determines the physical configuration of these channels (e.g. optical fibers) and measurement devices associated with that link. Possible protocols include the cavity-mediated Duan-Kimble protocol~\cite{Duan_2004} and the photon emission-based Barrett-Kok~\cite{Barrett_2005} and Cabrillo~\cite{Cabrillo_1999} protocols. These latter protocols employ an intermediate detector station between nodes to herald successful entanglement. 
\textsf{\textbf{(b)}}~Components in a repeater node. We classify repeater nodes based on the number of other nodes they connect to: a 4-way and a 3-way repeater are labeled. All adjacent repeater nodes are connected by quantum and classical channels (shown explicitly in black and light red respectively). 
\textsf{\textbf{(c)}}~Detector station used in the photon emission-based protocols to herald entanglement.
} \label{fig:q-network}
\end{figure*}

Figure~\ref{fig:q-network}a shows a schematic of such a repeater-connected quantum network.
A graph of quantum repeaters connected by quantum links forms the backbone of the network. Client nodes are quantum computers that connect to the network through their nearest repeater node, while repeater nodes facilitate the sharing of entanglement between clients. Hidden under the link layer abstraction~\cite{Dahlberg_2019} lies a physical layer of repeater devices and lossy quantum channels, the components of which are illustrated in the Figure. A typical repeater device, shown in Fig.~\ref{fig:q-network}b, consists of a memory capable of generating and storing entanglement with a photonic mode. The entanglement scheme used in a link between two repeater nodes determines the devices and channels used in the physical layer.
Nitrogen vacancy (NV) centers in diamond have been entangled at a distance of 1.3~km~\cite{Hensen_2015} using the emission-based scheme proposed by Barrett and Kok~\cite{Barrett_2005}, which uses a 50:50 beamsplitter to erase which-path information and detectors to herald entanglement. A related scheme proposed by Cabrillo~\cite{Cabrillo_1999} was used to provide on-demand entanglement between NV centers using a phase-stabilized fiber link~\cite{Humphreys_2018}. Recently, the direct-transmission (PLOB) bound \cite{Pirandola_2017} for quantum communication was broken~\cite{Bhaskar_2020} using a scheme proposed by Duan and Kimble~\cite{Duan_2004}, which relies on scattering photons from spins in high-cooperativity cavities.
These protocols are all inherently probabilistic, such that a single attempt at distributing entanglement succeeds with probability \pdistant.

The latency due to two-way communication for entanglement distribution in first-generation repeater networks forms a bottleneck that can be resolved by multiplexing of many quantum memories at each repeater node~\cite{Muralidharan_2016}.
To overcome this problem of latency, Ref.~\cite{Munro_2010} introduced a multiplexing scheme that is restricted to cavity-based entanglement protocols.
Alternatively, Ref.~\cite{Dam_2017} proposed a scheme that is compatible with emission-based protocols, but does not maintain the fidelity of entanglement distribution.
Previous papers considered switching to improve quantum key distribution rates, but have not quantified the effects of local connectivity on infidelities~\cite{Razavi_2009,Azuma_2015,Pirandola_2020}. Here, we introduce a quantum router scheme that uses multiplexing and all-to-all conditional local switching to increase entanglement fidelities and maintain entanglement rates for leading memory types and entanglement protocols. Our router architecture bears similarities to the quantum key distribution protocol introduced in a contemporaneous proposal by Tr\'{e}nyi and L\"{u}tkenhaus \cite{Trenyi_2020}, but our paper considers spin-photon implementations of quantum repeaters in general quantum networks.

The router uses local, low-loss connections to link different users' entangled qubits, thereby establishing entanglement across the channel. This quantum router architecture is motivated by recent advances in integrated photonics, with demonstrations of fast and low-loss on-chip switching~\cite{Desiatov_2019} of many photonic modes and the integration of quantum emitters with integrated circuits~\cite{Wan_2020}. 

In comparison with previous works, our proposed quantum router architecture is compatible with any entanglement generation protocol and quantum memory in the physical layer. Furthermore, our architecture extends naturally to chains of routers; in fact, we show that if brokered entanglement is used, the router can redirect entanglement flows to minimize latencies using only local information. We demonstrate these benefits by comparing the router architecture with a standard multiplexed architecture in which entanglement between users is generated serially~\cite{Rozpedek_2019,Pompili_2021}, exploring the various impacts on repeater performance.

\section*{Results}
\subsection*{Router Architecture}

We analyze the performance of our quantum router architecture for the NV center in diamond as the qubit platform and the Barrett-Kok~\cite{Barrett_2005} scheme as the entanglement protocol. However, our architecture is agnostic to the specific physical memory and entanglement-generation protocol.

\begin{figure*}[ht]
    \centering
    \includegraphics[scale = 1]{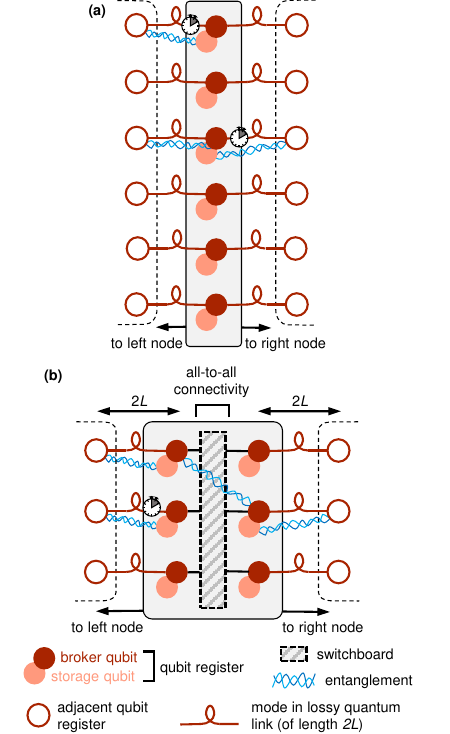}
    \caption{Abstract representation of a 2-way standard repeater and a 2-way router in a quantum network. Each repeater node hosts $m$ qubit registers, each of which here is formed by an NV center in diamond comprising a optically active electron spin (dark red) coupled to a long-coherence nuclear spin (light red), where the former serves as a broker to store entanglement in the latter. Multiple optical modes can be transmitted through the same physical link, e.g. via temporal or spectral multiplexing. \textbf{\textsf{(a)}}~Standard routerless repeater. In a standard repeater, each qubit register has lossy quantum links to registers at adjacent network nodes on both the left and the right. \textbf{\textsf{(b)}}~Router. In our router architecture, each qubit register has lossy links to only one of the left or right neighbors; however, the repeater node contains local, low-loss quantum channels that enable entanglement generation between all registers in the left bank and all registers in the right bank. As a result, the mean time a storage qubit must idle $t_{\text{idle}}$ while holding entanglement is shorter with a router. The idling time depends on $t_\text{distant}$ (the network clock period), $t_\text{local}$ (the local clock period), $p_\text{distant}$ and $p_\text{local}$ (the probability of generating entanglement over a distant and local link respectively).}
    \vspace{-6pt}
\label{fig:abstract}
\end{figure*}

In Fig.~\ref{fig:abstract}, we show the connectivity between photonic and stationary qubits in (a) a standard multiplexed repeater considered for diamond color centers~\cite{Rozpedek_2019}, and (b) a multiplexed repeater with a router. 
In this representation, each qubit register within the router is linked by a distinct mode to a corresponding register at an adjacent neighboring node, here labelled left and right for generality.
Different modes are shown separately, though they may be transmitted through the same physical channel, for example via temporal or spectral multiplexing. Each qubit register comprises a single NV center with two physical qubits: one electron spin (dark red), which can be used for optically-mediated entanglement, and one nuclear spin (light red), which has no optical transition but which can use spin-spin interactions to couple to and store quantum information from its local electron spin. This enables the use of `brokered entanglement'~\cite{Benjamin_2006}, where the electron spin serves as a short-term `broker' of entanglement to the longer-term `storage' qubit of the nuclear spin.

In the routerless architecture (Fig.~\ref{fig:abstract}a), each of the $m$ qubit registers first establishes entanglement with the left neighboring node via its electron spin and the corresponding optical mode. Letting $p_\text{distant}$ be the success probability of one entanglement generation attempt, the process takes an average number of attempts $1/p_\text{distant}$. Once successful, this entanglement is swapped to the nuclear spin for storage, at which point the electron spin attempts to establish entanglement with the right neighbor, again requiring an average of $1/p_\text{distant}$ attempts. Thus, the nuclear spin must idle while storing entanglement for an average time $t_{\text{idle}}\sim t_\text{distant}/p_\text{distant}$, where $t_\text{distant}$ is the time needed per attempt of the entanglement protocol, typically limited by the round-trip communication time between the repeater and a detector station. Once this succeeds, a Bell state measurement (BSM) on the joint electron-nuclear spin state teleports entanglement to be shared between the left and right nodes. This protocol is performed simultaneously and independently on each qubit register.

In contrast, our router architecture (Fig.~\ref{fig:abstract}b) defines two `banks' of registers, one with $\frac{m}{2}$ registers with optical links to the left, the other with $\frac{m}{2}$ registers with optical links to the right. The router connects these two banks using a low-loss $\frac{m}{2}\times\frac{m}{2}$ switchboard, over which entanglement protocols succeed with probability $p_\text{local} \gg p_\text{distant}$. In this architecture, each qubit register first establishes entanglement with either the left or the right, as determined by their optical connectivity. As in the routerless case, each link requires an average of $1/p_\text{distant}$ attempts before succeeding, after which these `successful' registers will then swap entanglement to their nuclear spin. At this point, any successful registers enter a `pairing' stage where registers in opposite banks are paired up and subsequent clock cycles attempt entanglement between electron spins in these pairs. This requires an average number of attempts $1/p_\text{local}$, taking a time $t_{\text{idle}}\sim t_\text{local}/p_\text{local}$; here, the local clock period $t_\text{local}$ is likely dominated by any state initialization required by the entanglement protocol. Any successful but unpaired registers idle, waiting for a successful partner on the opposite side to become available. Once entanglement is formed within a pair, two electron-nuclear spin BSM's --- one at each register --- teleport entanglement to be shared across the full left-right link.

\begin{figure*}[ht]
    \centering
    \includegraphics[scale = 0.4]{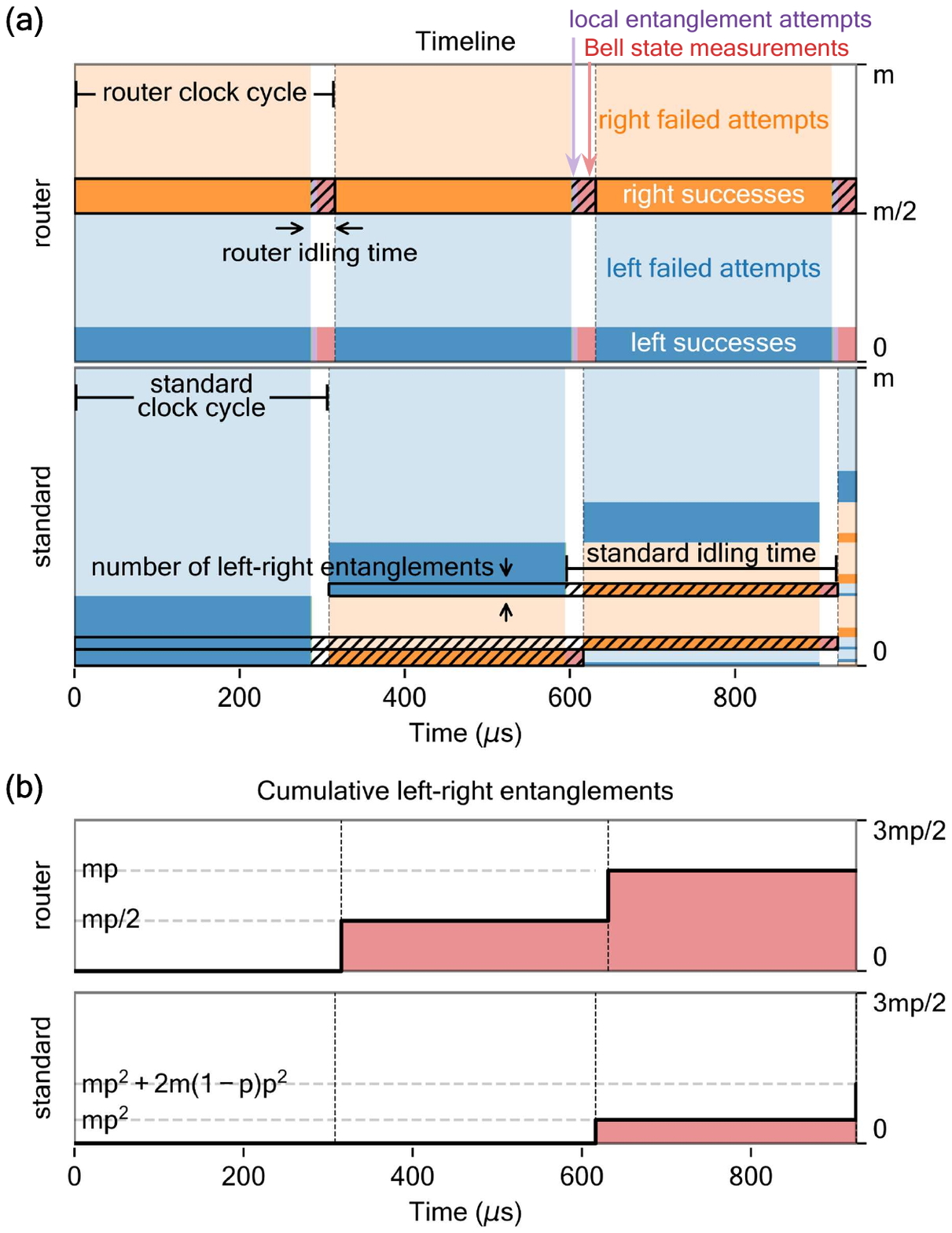}
    \caption{Timeline of activity for each memory in a standard repeater and a router, in the limit of high multiplexing ($m \rightarrow \infty$). Note that for visibility, here $p_{distant}$ is set to 0.2, which is 250$\times$ the value in our simulations. \textbf{\textsf{(a)}}~The horizontal axis represents time -- three router clock cycles are shown here -- and the vertical axis represents the number of memories involved in each stage of entanglement generation. The different stages of entanglement generation are color-coded. Full left-right entanglements are represented by boxes, the height of which is the number of left-right entanglements and the length of which is the time from the first successful distant entanglement attempt to the formation of left-right entanglement. The idling time of the first distant entanglement is hatched. Both the standard repeater and router are initialized with no entanglement at \mbox{time = 0 $\mu \text{s}$}. The router has a significantly shorter idling time $t_\text{idle}$ than the standard repeater.
    (See Supplementary Note~5 for more details.) \textbf{\textsf{(b)}}~The horizontal axis represents time (on the same scale as in \textbf{\textsf{(a)}}), but the vertical axis represents the cumulative number of Alice-Bob entanglements generated at any given time.
    }
    \vspace{-6pt}
\label{fig:timeline}
\end{figure*}

Figure~\ref{fig:timeline} compares the timeline of activity for a router and a standard repeater. Even though the router requires an additional `pairing' stage, this stage requires a negligible amount of time compared to the time needed for a distant entanglement attempt $t_\text{distant}$, so the router and the standard repeater can have similar clock cycles. Moreover, the router uses the entanglement with its neighboring nodes more quickly than the standard repeater does, so the router reaches its steady-state rate in a smaller number of clock cycles than the standard repeater.

This router architecture has two primary advantages even at steady state.
First, the idling time $t_\text{idle}$ that a storage qubit must hold entanglement can be greatly reduced, decreasing errors due to decoherence. Second, the number of entanglement attempts a register must make while its storage qubit holds entanglement is reduced by a factor $p_\text{local}/p_\text{distant}$, thus sharply reducing the storage-qubit decoherence due to broker-qubit entanglement attempts. In the remainder of this article, we quantitatively examine the impact the router has on the steady-state performance of a quantum network.

\subsection*{Simulations}
We use the NetSquid discrete event simulator~\cite{Coopmans_2021} to compare the performance of repeaters with and without a router, computing the average rate and fidelity of entanglement distribution for a one-repeater network. In this network, a single repeater station connects Alice and Bob, as shown in Fig.~\ref{fig:abstract} where Alice is the `left node' and Bob is the `right node'. We compare each architecture using the same number of total NV qubit registers $m$ at the repeater, under the assumption that qubit registers will be a scarce resource in near-term quantum networks. We choose the emission-based Barrett-Kok protocol~\cite{Barrett_2005} for our entanglement protocol. The physical parameters of our repeater memories correspond to a realization using photonic integrated circuits discussed below, using experimentally-realized values reported in the literature (see Supplementary Note~1 for details).

Figure~\ref{fig:results}b plots the entanglement distribution rates for both repeater architectures for various link distances as a function of the number of qubit registers $m$ at the repeater node. For the routerless case, the rate scales linearly with $m$ since the qubit registers operate independently. For low $m$, the router exhibits comparatively lower rates; however, as $m$ increases, the difference between the two protocols decreases, such that in the limit of large $m$ the two architectures perform comparably. 
This difference can be attributed to mismatches in the number of successful entanglements in Alice’s and Bob’s banks after a given clock cycle. The delay in resolving these mismatches, the number of which scales with $\sqrt{m}$, lowers the rate of the router below the linear scaling of the routerless architecture. However, the fractional impact of this effect on the rate is reduced for large $m$, and the rate of the router approaches the routerless rate (see Supplementary Note~2 for a lower bound on the router rate).

\begin{figure*}[ht!]
    \centering
    \includegraphics[scale = 1]{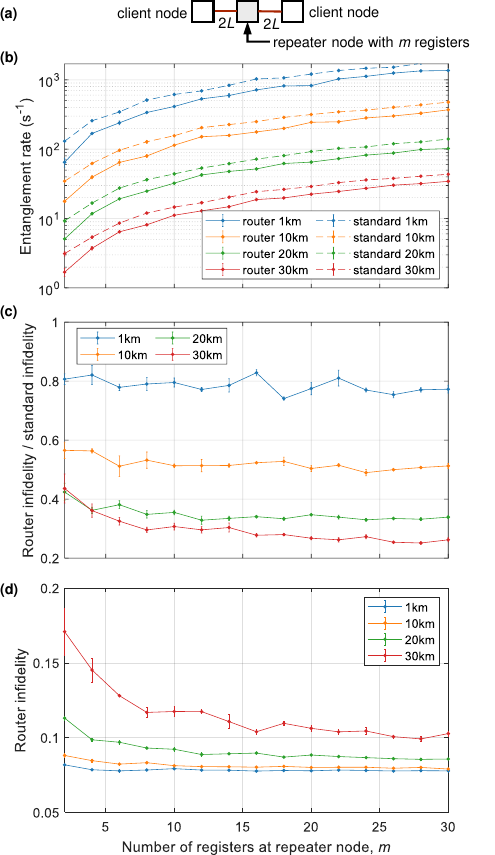}
    \vspace{-6pt}
    \caption{Simulation results for different repeater configurations. Each data point is the result of 3 independent simulations. The average rates and infidelities are plotted, with the standard error of the mean shown as the error bars. The lengths (1km, 10km, 20km, 30km) refer to the distance $L$ between the repeater node and the detector station. \textbf{\textsf{(a)}}~Link layer representation of the simulated one-repeater network. \textbf{\textsf{(b)}}~Steady-state entanglement rate against $m$. As expected, the rate of entanglement generation increases with the number of repeater qubit registers ($m$), as expected. The entanglement rate of the router architecture is slightly lower than the entanglement rate of the standard repeater, and the difference decreases as the size of the repeater increases. \textbf{\textsf{(c)}}~Ratio of entanglement infidelity against $m$. The fidelity of entanglement generated by the router is higher than that of the standard repeater for sufficiently large distances $L$. The fidelity of entanglement generated by the router improves as $m$ increases, whereas that of the standard repeater does not. This causes the infidelity ratio to decrease as $m$ increases. \textbf{\textsf{(d)}}~Router infidelity against $m$. As $m$ increases, the router infidelity approaches a channel-loss-invariant value that only depends on local gate fidelities and the round-trip travel time.}
    \label{fig:results}
\end{figure*}

Figures~\ref{fig:results}c-d plot the infidelity of the distributed entanglement for repeaters with and without a router. We consider three sources of infidelities in the distributed entanglement. The first source of infidelity is the typical depolarizing and dephasing noise experienced independently by the electron and nuclear spins, as characterized by their $T_1$ and $T_2$ coherence times. We model this noise on a qubit after time $t$ as:
\begin{equation}
    \rho = \begin{pmatrix}
    1-\rho_{11} & \rho_{01} \\ \rho^*_{01} & \rho_{11}
    \end{pmatrix} \mapsto \begin{pmatrix}
    1-\rho_{11}e^{-t/T_1} & \rho_{01}e^{-t/T_2} \\ \rho^*_{01} e^{-t/T_2} & \rho_{11} e^{-t/T_1}
    \end{pmatrix}.
\end{equation}

For a repeater with no router, adding additional registers does not affect the amount of time an individual register must store entanglement; thus, increased multiplexing does not reduce this decoherence channel. However, for a repeater with a router, the number of registers has a dramatic effect on the fidelity. This observation can be understood by considering the mean idling time for a memory in this scheme, which is determined by the number of clock cycles a given register in the Alice (Bob) bank must idle after establishing entanglement on its respective side of the link before a register in the Bob (Alice) bank also succeeds and is available for the pairing stage of the protocol. This maps to the success mismatch problem discussed earlier; as such, in the limit of large $m$, most successful registers will be paired with a register in the opposite bank with every clock cycle, such that the idling time approaches the duration of a single such cycle. In contrast, with no router, a register must idle for on average $1/p_{\text{distant}} \gg 1$ cycles. Thus, the addition of a router reduces the infidelity from this channel by a factor that approaches \pdistant. 

The second source of error stems from coupling between the nitrogen nuclear spin and the NV electron spin. In particular, each excitation of the electron spin used to generate spin-photon entanglement in the Barrett-Kok protocol generates noise that decoheres the qubit stored in the nuclear spin. We can model this interaction-induced noise on the nuclear spin as \cite{Rozpedek_2019}:
\begin{equation}
    \rho \mapsto (1-a-b)\rho + aZ\rho Z + b \frac{ \mathbb{I}}{2},
\end{equation}
where the dephasing $a \approx 1/4000$ and the depolarization $b \approx 1/5000$ for NV centers. Note that for other systems with stronger electron-nuclear spin coupling such as silicon vacancy centers in diamond~\cite{Nguyen_2019}, this effect may be stronger. For both architectures, after a register has generated one entanglement link, it stores that entanglement in the nuclear spin while its electron spin attempts to generate entanglement on the other side --- either directly to Bob in the case with no router, or to the opposite bank when using a router. In either case, each failed attempt decreases the fidelity of the final Alice-Bob EPR pair. Without a router, the mean number of failed attempts is equal to $1/p_{\text{distant}}$; with a router, this number is instead $1/p_{\text{local}}$, dramatically reducing the infidelity from this effect.

The third source of infidelity comes from imperfect two-qubit gates and electron spin readout. The proposed router needs to perform twice as many two-qubit gates and readouts for each distributed EPR pair as a standard repeater. The cumulative infidelity of these operations is independent of $m$, \pdistant{}, and \plocal{}. When the link between adjacent nodes is short, these operations dominate the infidelity of the Alice-Bob EPR pair, disadvantaging the router. However, when the link becomes lossier, the previous two sources of infidelity dominate, causing the router to increasingly outperform the standard repeater in entanglement fidelity (see Supplementary Note~3 for more details).

As a result, the router enables channel-loss-invariant entanglement fidelity; that is, regardless of the network link efficiency, the router can achieve fidelity that is limited only by the gates performed on the quantum memories and decoherence during a single round-trip transit of the network. 
A threshold is crossed as the number of entanglement attempts per clock cycle exceeds $1/p_{\text{distant}}$, where the average clock cycle produces at least one entanglement event on both sides of the router. This limits the decoherence a memory must endure to that of one clock cycle, and, for the case where the memory coherence time is much longer than the link round-trip time, allows all routers to realize a fidelity that is purely limited by the fundamental gates necessary to perform the protocol (initialization, swapping, measurement, etc.). This asymptotic channel-loss-invariant fidelity is illustrated in Fig.~\ref{fig:results}d. While increasing the distance between adjacent nodes increases the round-trip time and thus the decoherence in one clock cycle, Fig.~\ref{fig:results}c shows that the asymptotic infidelity increases less than it would have if it were to depend on the channel loss as well.

\section*{Discussion}

\begin{figure*}[ht!]
    \centering
    \includegraphics[scale = 1]{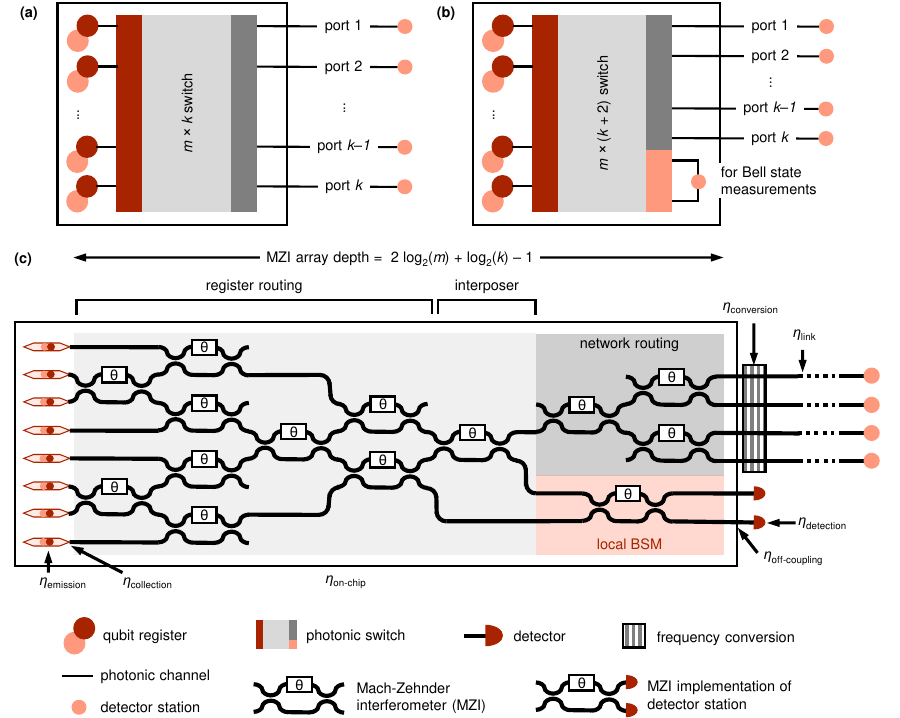}
    \caption{Potential physical realization of $k$-way repeater nodes with $m$ repeater qubit registers. The main component of our proposed router architecture is a photonic switch that provides all-to-all connectivity from ports on one side of the switch to ports on the other side. Photonic switches can take the form of a tree of Mach-Zehnder interferometers (MZIs) when the modes (Fig.~\ref{fig:abstract}) are temporally multiplexed, or a wavelength-division multiplexer when the modes are spectrally multiplexed. \textbf{\textsf{(a)}}~Standard repeater realization. A standard (routerless) repeater can be constructed using an $m \times k$ switch that connects all $m$ repeater registers with the $k$ links that go to other nodes in the network. \textbf{\textsf{(b)}}~Router realization. A router can be constructed using an $m \times (k+2)$ switch. Photons emitted by the $m$ repeater registers can travel to the $k$ links that go to other nodes in the network, or they can be sent to a local detector station for local entanglement generation. \textbf{\textsf{(c)}}~MZI implementation of $k=4$-way, $m=8$-register router. A $k$-way, $m$-register router can be implemented using an MZI array of depth $\log_2(m^2k/2)$, comprising a register routing layer (depth $2\log_2(m/2)$), an interposer for routing from local to distant connections (depth $1$), a network routing layer (depth $\log_2(k)$), and a local BSM layer (depth 1). The ends of the MZI array are coupled either directly to detectors for the local BSMs or a frequency conversion stage followed by fiber links to distant detector stations for networking. All losses considered in our simulations are shown here.}
    \label{fig:physical}
\end{figure*}

Though the complete bipartite connectivity of routers may appear challenging to implement, a router can be added to a standard multiplexed repeater with relatively little physical resource overhead. Figure \ref{fig:physical} illustrates one such realization using photonic integrated circuits (PICs). Quantum registers are integrated onto PICs that contain an array of Mach-Zehnder interferometers (MZIs)~\cite{Harris_2017}, forming a fast switching network to connect any register to any output channel. For a standard multiplexed $k$-way repeater with no local connectivity, this requires an $m\times k$ switch to connect $m$ repeaters to any of the possible $k$ neighbors in the network, as shown in Fig.~\ref{fig:physical}a. However, a router can be embedded in this architecture by extending this to an $m\times (k+2)$ switch. The additional two ports lead to a detector station as shown in Fig.~\ref{fig:physical}b, thus enabling photon-mediated entanglement between any two registers in the repeater. Since typical single photon detectors have dead times of a few tens of nanoseconds, a single pair of detectors could facilitate many attempts at local entanglement in a single repeater clock cycle, which may be hundreds of microseconds long depending on the length of the link being connected. The additional complexity involved in adding these two additional channels is small, making this approach an attractive option for implementing all-to-all local connectivity. 

On a PIC, the $m\times k$ switch required for the routerless repeater can be realized on an MZI array of depth $\log_2(m)+\log_2(k)$~\cite{Lee_2019}. However, this architecture does not permit simultaneous routing from multiple memories, which is desired for the local entanglement operations used in the quantum router. In Fig.~\ref{fig:physical}c we present an architecture that permits arbitrary $m\times k$ routing for connecting registers to the network, allows simultaneous routing from two memories to perform Bell state measurements for local entanglement generation, and maintains $\mathcal{O}(\log m)$ scaling of the array depth. The addition of a one-MZI interposer permits routing between the network and a Bell state measurement setup for local entanglement generation (see see Supplementary Note~4 for more details). The simulations presented in Fig.~\ref{fig:results} utilize this architecture assuming an aluminum nitride (AlN)-based photonic chip; alternative implementations include using silicon nitride (SiN)~\cite{Mouradian_2015} or lithium niobate (LN)~\cite{Desiatov_2019}.
The control requirements for such large MZI arrays are similar to those required by and demonstrated for optical neural nets~\cite{Shen_2017} and programmable photonic circuits~\cite{Harris_2018,Bogaerts_2020}, allowing us to take advantage of this existing technology for our router. 
Additional improvements include integrating detectors on-chip~\cite{Najafi_2015} to eliminate the off-chip coupling inefficiency for local measurements and integrating the frequency conversion stage on-chip, which is feasible in highly nonlinear materials such as LN~\cite{Wang_2018}.

This platform is an attractive path for realizing recent proposals for on-chip polarization-to-spin based networking~\cite{Chen_2020}. The PIC implementation presents an additional benefit if the links to Alice and Bob have different transmissivities. Since both the rate and fidelity of entanglement distribution are limited by the difference in the time needed to establish entanglement with both Alice and Bob, the repeater in Fig.~\ref{fig:abstract}b (with equal numbers of registers in the Alice and Bob banks) will be limited by the lossier side. In contrast, the PIC implementation allows dynamic allocation of registers to Alice and Bob to balance the rates of entanglement generation. This optimizes the entanglement rate by effectively increasing the multiplexing on lossier ports, and optimizes the entanglement fidelity by preventing a build-up of idling registers on a lower-loss port. Dynamic memory allocation is also of great importance for managing entanglement flows in general 2D quantum network geometries~\cite{Pant_2019,Pirandola_2019,Shi_2020}.

The router architecture has additional advantages in multi-repeater quantum networks. The high local connectivity of the router enable chains of repeaters to automatically connect the longest available entanglement links, favoring the generation of a single EPR pair shared between distant nodes over that of multiple EPR pairs shared between relatively proximate nodes.
Specifically, the first-in, first-out behavior of a router guarantees that entanglement will always grow from the longest established chain. 
In contrast, even if we allow adjacent standard repeaters to have all-to-all connectivity between registers in adjacent nodes (e.g. by spectral shifting or time bin reordering), there is no guarantee that the longest links will be generated first.

\begin{figure*}[ht!]
    \centering
    \includegraphics[width = 0.45\textwidth]{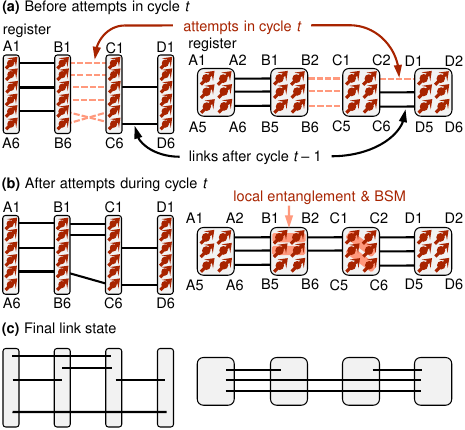}
    \caption{Dynamics of repeater chains. Four repeater nodes in each chain are shown for both the standard repeater chain (left panels) and the router chain (right panels). Both chains have the same number of successful entanglement links. Registers in each node are labeled from A1 to D6. \textbf{\textsf{(a)}}~State of the chain before the entanglement attempts during some clock cycle $t$. The entanglement attempts to be made in clock cycle $t$ are shown as dashed lines. \textbf{\textsf{(b)}}~State of the chain after the entanglement attempts during clock cycle $t$, but before Bell state measurements are performed. Bell state measurements in the router chain are performed according to the first-in-first-out strategy. \textbf{\textsf{(c)}}~Link state after Bell state measurements during clock cycle $t$. The standard repeater chain has one long link that spans all four nodes, but the router chain has two long links. The standard repeater chain is unable to connect the shorter links that are available to form long links.}
    \label{fig:chaining}
\end{figure*}

This behavior is illustrated in Fig.~\ref{fig:chaining}, which compares the dynamics of a standard network and a router network under similar scenarios. Before each clock cycle, a repeater or router will determine which links to attempt on the subsequent clock cycle given the current link state. An example is shown in Fig.~\ref{fig:chaining}a.
If we further allow the standard repeater to have knowledge of the global network state, it can adjust its transmission sequence to prioritize a longer link, e.g. by attempting to entangle registers B5 and C6 (instead of B5-C5 and B6-C6). In contrast, the router does not require any information about the network state; each register simply attempts entanglement with a fixed partner in a neighboring node.

Figure~\ref{fig:chaining}b shows a possible network state after the entanglement attempts in a clock cycle. Each router uses local entanglement and Bell state measurements to connect its left and right banks of registers. Following a first-in-first-out strategy, it connects the first available entangled register in the left bank with the first available entangled register in the right bank, and repeats until only unentangled registers remain in one of the banks. Using only local information, the router network thus automatically connects as many long links as possible, as demonstrated by the link state in Fig.~\ref{fig:chaining}c. On the other hand, the standard repeater may not be able to connect long links even if there are enough successful entanglements to do so. As a result, the state of the standard network after this time step yields some long links, but also many singular short links that cannot be merged.

The router's strategy reduces the idling time between local entanglement generation and the delivery of this entanglement to the end users, thereby increasing the fidelity of the delivered entanglement. We highlight that this router behavior is realized without any need for knowledge of the global network state. In a true deployed network, communication latencies prevent individual nodes from knowing the state of the total network in real-time, making the router's ability to optimize fidelity with local-only information all the more notable.

In conclusion, we have addressed a critical problem in quantum repeater architectures: minimizing memory latencies to maintain entanglement fidelity across repeater networks. Specifically, the quantum router architecture enables entanglement fidelity that is invariant to the channel losses in the network. Our architecture applies to leading entanglement protocols and automatically prioritizes entanglement flows using only local information. An important next step will be to exploit the local connectivity for entanglement distillation~\cite{Kalb_2017} and local error correction~\cite{Cramer_2016} to further improve the fidelity of generated entanglement. In this way, the entanglement rate can be traded for entanglement fidelity, without modification to the router architecture. 
While the above analysis considered NV centers in diamond, the architectural benefits should apply to other quantum memory modalities, including solid-state artificial atoms, trapped ions and neutral atoms.
In suitable parameter regimes, routers can also incorporate multiplexing into realizations of distance-independent entanglement rates using local GHZ projective measurements~\cite{Patil_2020}.
Our results emphasize the importance of local connectivity in designing multiplexed quantum repeaters for high-rate and high-fidelity entanglement distribution across quantum networks.

\section*{Data availability}
The data that support the plots within this paper and other findings of this study are available from the corresponding author upon reasonable request.

\section*{Code Availability}
The NetSquid simulation package is available for download by request at https://netsquid.org/. The additional code used to generate data in this work is presently available at https://github.com/leeyuan13/NetSquidSim.

\section*{Acknowledgements}

Y.L. acknowledges funding support by the MIT School of Engineering's SuperUROP program. E.B. was supported by a NASA Space Technology Research Fellowship and the NSF Center for Ultracold Atoms (CUA). A.D. and S.W. were supported by an ERC Starting grant, EU Flagship on Quantum Technologies, Quantum Internet Alliance, an NWO VIDI grant, and the Zwaartekracht QSC. D.E. acknowledges support from the NSF EFRI-ACQUIRE program Scalable Quantum Communications with Error-Corrected Semiconductor Qubits and the NSF ERC Center for Quantum Networks (CQN). We also thank Ian Christen, Kevin Chen, and Robert Knegjens for valuable discussions. 

\section*{Author contributions}
Y.L. and E.B. designed and performed the simulations. D.E. conceived the idea and supervised the project. A.D. and S.W. developed the code framework to run simulations using NetSquid. All authors contributed to writing and revising the manuscript.

\section*{Competing Interests}
The authors declare no competing financial or non-financial interests.

\bibliography{references_limited}

\clearpage

\end{document}


\title{Supplementary Information for ``A Quantum Router Architecture for High-Fidelity Entanglement Flows in Quantum Networks''}

\author{Yuan Lee}
\affiliation{Department of Electrical Engineering and Computer Science, Massachusetts Institute of Technology, Cambridge, Massachusetts 02139, USA}

\author{Eric Bersin}
\affiliation{Department of Electrical Engineering and Computer Science, Massachusetts Institute of Technology, Cambridge, Massachusetts 02139, USA}

\author{Axel Dahlberg}
\affiliation{QuTech, Delft University of Technology, and Kavli Institute of Nanoscience
Delft, The Netherlands}

\author{Stephanie Wehner}
\affiliation{QuTech, Delft University of Technology, and Kavli Institute of Nanoscience
Delft, The Netherlands}

\author{Dirk Englund}
\affiliation{Department of Electrical Engineering and Computer Science, Massachusetts Institute of Technology, Cambridge, Massachusetts 02139, USA}
\affiliation{Research Laboratory of Electronics, Massachusetts Institute of Technology, Cambridge, Massachusetts 02139, USA}

\maketitle

\appendix
\renewcommand{\appendixname}{Supplementary Note}
\renewcommand{\thesection}{\arabic{section}}
\counterwithout{equation}{section}

\section{Simulation Parameters}
\label{app:parameters}

The physical parameters of our repeater memories correspond to experimentally-realized values reported in literature, detailed in Supplementary Table~\ref{supptab:parameters}. In particular, these simulations consider a number of loss mechanisms. Only a fraction $\eta_{\text{emission}}$ of the NV emissions are coherent and thus usable for quantum networking, with the remainder being scattered into an incoherent phonon-sideband. A common strategy to improve this fraction is to place the NV in a cavity which enhances coherent emissions via the Purcell effect. Our simulations assume such a strategy, using the best-demonstrated enhancement with NV centers~\cite[][]{Li_2015}. The NV also emits into a wide range of spatial modes, yielding a limited collection efficiency $\eta_{\text{collection}}$. Local routing at the repeater is performed with efficiency $\eta_{\text{local}}$, determined by the loss of the photonic material, here assumed to be a recently-demonstrated aluminum nitride (AlN) platform. The NV's visible emission wavelength experiences high losses of around 8 dB/km in fiber~\cite[][]{Hensen_2015}; as a result, long-distance networking applications require a frequency downconversion step to telecom wavelengths with efficiency $\eta_{\text{conversion}}$. The primary source of loss is scattering in the optical fiber that links the network nodes. This loss scales exponentially with the link length $L$, giving transmission efficiency $\eta_{\text{link}}=e^{-\alpha L}$. The transmission losses $\eta_\text{transmission}$ labeled in 
Fig.~5
combine contributions from frequency downconversion and scattering losses in the fiber. Finally, the single photon detectors used to herald entanglement have a finite detection efficiency $\eta_{\text{detection}}$ and dark count rate $d$.

There are two components to the efficiency of local routing $\eta_\text{local}$: off-chip coupling efficiencies ${\eta_\text{off-coupling} = 0.92}$ \cite[][]{Notaros_2016} and on-chip propagation efficiencies ${\eta_\text{on-chip} = 0.9963}$ per MZI. To compute the latter, we assume that AlN on sapphire waveguides have a loss of 5.3 dB/cm with a footprint of 30 \textmu m per MZI~\cite[][]{Lu_2018}.

\renewcommand{\arraystretch}{1.1}
\begin{table*}
\caption{State-of-the-art parameters used in simulations. Noise from two-qubit gates and electron spin readout are modeled as depolarization; noise from memory-photon state preparation is modeled as dephasing.} \label{supptab:parameters}
\setlength\tabcolsep{3pt}
\begin{tabularx}{0.8\textwidth} { 
   >{\raggedright\arraybackslash}c 
   >{\raggedright\arraybackslash}c 
   >{\raggedright\arraybackslash}l }
 \hline
 Parameter & Value 
                            & Description \\ \hline \hline
 $\eta_{\text{emission}}$ & 0.66~\cite[][]{Li_2015} & Fraction of NV emissions that are coherent\\ \hline 
 $\eta_{\text{collection}}$ & 0.83~\cite[][]{Mouradian_2015} & Collection efficiency from NVs \\ \hline 
 $\eta_{\text{local}}$ & (Refer to text.) & Efficiency of local routing \\ \hline
 $\eta_\text{conversion}$ &  0.33~\cite[][]{Yu_2020} & Efficiency of frequency conversion \\ \hline
 $\eta_\text{detection}$ & 
    0.93~\cite[][]{Le_Jeannic_2016}
    & Detector efficiency \\ \hline
 $d$ & 
    $10 \, \text{s}^{-1}$
    & Detector dark count rate \\ \hline
 $T_1^\text{e}$ & 3600 s~\cite[][]{Abobeih_2018} & Electron depolarizing time \\ \hline 
 $T_2^\text{e}$ & 1.58 s~\cite[][]{Abobeih_2018} & Electron dephasing time \\ \hline 
 $T_1^\text{n}$ & $> 86400 \text{s}$  & Nuclear depolarizing time (too long to measure accurately) \\ \hline 
 $T_2^\text{n}$ & 63 s~\cite[][]{Bradley_2019} & Nuclear dephasing time \\ \hline
 $a$ & 1/4000~\cite[][]{Reiserer_2016, Kalb_2018} & Interaction-induced dephasing \\ \hline
 $b$ & 1/5000~\cite[][]{Reiserer_2016} & Interaction-induced depolarization \\ \hline
$\alpha$ & 0.041 km$^{-1}$ & Fiber loss (0.18 dB/km  at telecom wavelengths) \\ \hline 
$F_\text{gate}$ & 
    0.992~\cite[][]{Rong_2015}
    & Fidelity of two-qubit gates  \\ \hline
$F_\text{readout}$ & 
    0.9998~\cite[][]{Bhaskar_2020} 
    & Readout fidelity from electron spin \\ \hline
$F_\text{initialization}$ & 
    0.99~\cite[][]{Hensen_2015} 
    & Fidelity of memory-photon state preparation \\ \hline
\end{tabularx}
\end{table*}

\section{Rates of One-Repeater Networks} \label{app:rate}
We can obtain simple expressions for the raw entanglement rate of single repeaters, not accounting for fidelities. In a standard routerless repeater, each register attempts entanglement on either the left link or the right link in a single network clock cycle $t_\text{distant}$. The mean number of successful long-distance links in a clock cycle is $mp_\text{distant}$, where $m$ is the number of registers per node and $p_\text{distant}$ is the corresponding probability of success (as described in the main text). Two successful long-distance entanglements (one on the left and one of the right) are needed to generate an EPR pair, so the mean entanglement rate for a standard repeater is $mp_\text{distant}/2t_\text{distant}$.

For the described router architecture, however, any given register only attempts entanglement on one of the links. We assume that $t_\text{local} \ll t_\text{distant}$, so that with optimized scheduling, the router can use the same clock cycle as the standard repeater. For our simulations, the time necessary for local entanglement formation is included. At the start of any clock cycle, some registers might be holding on to long-distance entanglement from the previous clock cycle, waiting to be matched with a new successful entanglement link from the other link. 

Let $\Delta_t$ be the signed number of such registers at the start of clock cycle $t$, so that $\Delta_t$ is the number of excess entanglements over the left link if $\Delta_t > 0$ and $-\Delta_t$ is the number of excess entanglements over the right link if $\Delta_t < 0$. Note that if there are excess entanglements over the left link, this means that all entanglements over the right link were matched in the previous clock cycle, and vice versa. Then $\lbrace \Delta_t \rbrace_{t=1}^\infty$ is a Markov chain with update law $\Delta_{t+1} = \Delta_t + B_\ell - B_r$, where
\begin{gather}
    B_\ell \sim \begin{cases}
    \text{Bino}(m/2-\Delta_t, p_\text{distant}) & \text{if $\Delta_t \geq 0$} \\
    \text{Bino}(m/2, p_\text{distant}) & \text{if $\Delta_t < 0$}
    \end{cases} \nonumber \\
    B_r \sim \begin{cases}
    \text{Bino}(m/2, p_\text{distant}) & \text{if $\Delta_t \geq 0$} \\
    \text{Bino}(m/2+\Delta_t, p_\text{distant}) & \text{if $\Delta_t < 0$}
    \end{cases}
\end{gather}
are independent binomial random variables representing the numbers of successful entanglement attempts on the left and right links respectively. ($B_\ell$ and $B_r$ have an implicit time dependence.) The number of registers with entanglements from the previous clock cycle is $\lvert \Delta_t \rvert$, so the number of new entanglement attempts in clock cycle $t$ is $m - \lvert \Delta_t \rvert$. Every pair of matched entanglements will establish local entanglement with high probability, allowing the router to perform the required Bell state measurements and generate an EPR pair across the repeater. Using the ergodic theorem, the mean entanglement rate for the described router architecture is hence $(m - \mathbb{E}\lvert \Delta_t \rvert ) p_\text{distant} / 2 t_\text{distant}$, where the expectation is taken over the stationary distribution of the Markov chain.

It is difficult to find $\mathbb{E}\lvert \Delta_t \rvert$ analytically, but we can bound it from above:
\begin{align}
    \mathbb{E} \lvert \Delta_t \rvert  &\leq \sqrt{\mathbb{E} \Delta_t^2} \text{ (Cauchy-Schwarz)} \nonumber \\
    &= \sqrt{\frac{m(1-p_\text{distant})}{2-p_\text{distant}}} \leq \sqrt{\frac{m}{2}}.
\end{align}
Hence, the router rate is at least $(1-1/\sqrt{2m}) mp_\text{distant} / 2t_\text{distant}$. This lower bound is close to the actual rates observed in the simulation (Fig.~4b).

With some modifications to the architecture, we can improve the router rate to be closer to the standard repeater rate. The router architecture described in the main text has $\mathbb{E}[\Delta_{t+1} \, \vert \, \Delta_t] = \Delta_t (1-p_\text{distant})$, causing imbalances in entanglements between the left and right links to persist over time. However, using the MZI implementation of the router (Fig.~5) allows the node to reassign registers to different banks in different clock cycles. By assigning different numbers of `unentangled' registers to the left and right banks at different clock cycles, we can choose $B_\ell, B_r$ such that $\mathbb{E}[\Delta_{t+1} \, \lvert \Delta_t] = 0$. Then $\mathbb{E}\lvert \Delta_t \rvert \leq \sqrt{mp_\text{distant}}$, so the router rate can improve to $(1-\sqrt{p_\text{distant}/m}) mp_\text{distant} / 2t_\text{distant}$. However, these modifications do not change the asymptotic behavior (i.e. $(1-\mathcal{O}(1/\sqrt{m}))mp_\text{distant}/2t_\text{distant}$) of the router rate.

\section{Fidelity Contributions} \label{app:fidelity}
Many factors contribute to the infidelity of entanglement generated by a router or a standard repeater. 
The impact of each factor on the total infidelity often depends on other factors, but we can still estimate their respective first-order contributions to the high fidelity of the router by mitigating individual factors and observing the resulting infidelity.

In the main text, we identified three contributions to the infidelity of entanglement. First, memories holding entangled states decohere during the idling time $T$ before they are paired with entanglement on the other side of the repeater. 
Second, each excitation of the electron spin used to produce spin-photon entanglement in the Barrett-Kok protocol generates interaction noise that disturbs the qubit stored in the nuclear spin. If the nuclear spin holds entanglement on one side of the repeater, the infidelity of Alice-Bob entanglement increases with the number of entanglement attempts $A$ made by the corresponding electron spin. 
Third, local operations like two-qubit gates and Bell state measurements further increase to the infidelity of entanglement.

\begin{figure}[ht!]
\centering
\includegraphics[width = 0.5\textwidth]{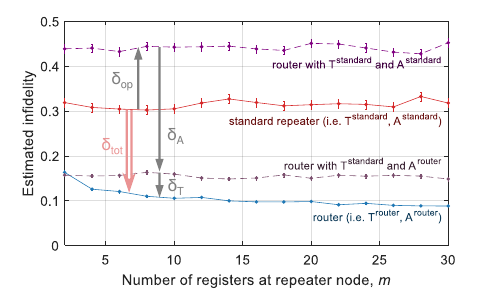}
\caption{Mean infidelity of entanglement generated by a one-repeater network. The distance between the repeater and the detector station (for the Barrett-Kok protocol) is $L = 30\text{km}$. $\delta_\text{tot}$ is the difference in fidelity between the standard repeater and the router; $\delta_\text{op}$ is the contribution from local operations; $\delta_A$ is the contribution from interaction noise; and $\delta_T$ is the contribution from decoherence. Comparing the standard repeater to the router, we estimate $(\delta_\text{op}, \delta_A, \delta_T) \approx (0.6 \delta_\text{tot}, 1.4 \delta_\text{tot}, 0.2 \delta_\text{tot})$. Error bars are standard errors of the mean.}
\label{suppfig:fidelitybreakdown}
\end{figure}

Figure~4 shows that the infidelity of the router is smaller than the infidelity of the standard repeater. As the router can simultaneously attempt entanglement with both Alice and Bob, its entangled states have a lower idling time than they would in the standard repeater, i.e. $\mathbb{E}T^\text{router} < \mathbb{E}T^\text{standard}$. Moreover, when the nuclear spin shares an entangled state with Alice or Bob, the electron spin requires fewer attempts to establish local entanglement in the router than it would to establish distant entanglement in the standard repeater, because the probability of successfully generating entanglement is higher over a low-loss local link than over a lossy distant link. In other words, $\mathbb{E}A^\text{router} < \mathbb{E}A^\text{standard}$. Both of these factors reduce the router infidelity relative to the standard infidelity.

However, the router performs more local operations than the standard repeater for each Alice-Bob EPR pair. The router needs to perform two Bell state measurements to connect the Alice-router, router-router and Bob-router entanglements, whereas the standard repeater only needs to perform one Bell state measurement per Alice-Bob EPR pair. The router thus experiences greater infidelity from local operations than the standard repeater. Nevertheless, the combination of local operations, interaction noise and decoherence times causes the router to have lower overall infidelity.

Let $\delta_\text{tot}$ be the difference between the router and standard infidelities. Let $\delta_\text{op}$ be the contribution to this difference from local operations, $\delta_A$ be the contribution from interaction noise (i.e. from the difference between the random variables $A^\text{router}$ and $A^\text{standard}$) and $\delta_T$ be the contribution from decoherence (i.e. from the difference between the random variables $T^\text{router}$ and $T^\text{standard}$). To first order, we assume that the infidelity contributions are not interdependent, so we should have
\begin{equation}
    \delta_\text{tot} = -\delta_\text{op} + \delta_A + \delta_T.
\end{equation}

To identify the relative contributions of these three factors to the reduction in router infidelity, we estimate the mean infidelity using a sampling approximation. For example, to estimate the mean router infidelity, we first generate Alice-router entanglement. We draw a sample of the idling time $T^\text{router}$ and apply the appropriate dephasing and depolarization operation to the Alice-router entanglement. Then we generate router-Bob entanglement, draw a sample of the number of local entanglement attempts needed $A^\text{router}$, apply the appropriate interaction noise to the Alice-router and router-Bob entanglements, and generate (local) router-router entanglement. Finally, we perform the two Bell state measurements, producing an Alice-Bob EPR pair. We repeat the procedure multiple times to generate samples of the Alice-Bob EPR pair, before approximating the mean router infidelity using the sample average.

This simplified simulation does not reproduce the results of the full NetSquid simulation (cf. main text) exactly, but we observe that the simplified simulations produce the same trends as the full simulation.

Using the same sampling approximation, we can also estimate the router infidelity in the counterfactual situation where $T^\text{router}$ and/or $A^\text{router}$ have the same distribution as $T^\text{standard}$ and/or $A^\text{standard}$ respectively. The estimated infidelities are shown in Supplementary Fig.~\ref{suppfig:fidelitybreakdown}.

Even though the contribution of local operations to the router infidelity is significant, with $\delta_\text{op} \approx 0.6 \delta_\text{tot}$, the reduction in infidelity due to reduced interaction noise is even greater, with $\delta_A \approx 1.4 \delta_\text{tot}$. We also observe that the decrease in router infidelity with increasing multiplexing arises from $T^\text{router}$. The average idling time of an entangled link with the router decreases as $m$ increases, leading to lower infidelity. In contrast, the other contributions are not affected by the degree of multiplexing.

In the limit as $m\rightarrow \infty$, the idling time of the router $T^\text{router}$ converges to zero, whereas the idling time of the standard repeater $T^\text{standard}$ is unchanged. As the channel loss affects $T^\text{standard}$ but not $T^\text{router}$ ($\xrightarrow{\text{p}} 0$) as $m\rightarrow \infty$, the router achieves channel-loss-invariant fidelity in the limit of high multiplexing.

Therefore, both the idling time and the number of entanglement attempts play important roles in reducing the infidelity of the router relative to that of the standard repeater.

\section{Design of MZI Array} \label{app:MZI}
\begin{figure}[ht]
\centering
\includegraphics[width = 0.5\textwidth]{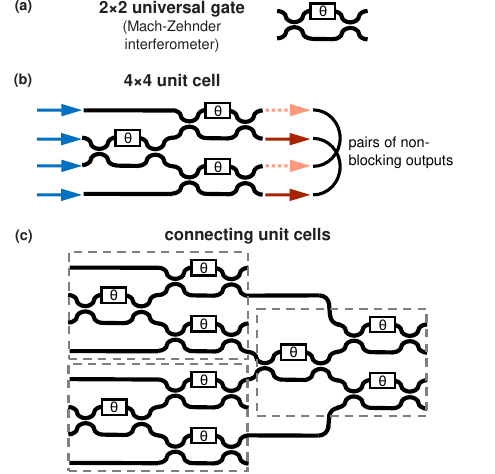}
\caption{Building the MZI array. \textbf{\textsf{(a)}}~The MZI implements $2\times 2$ universal gates, so by tuning the parameters of the MZI (e.g. the phase difference $\theta$) we can use the MZI to switch photons between adjacent waveguide lines. \textbf{\textsf{(b)}}~To achieve the non-blocking property at the two detector ports, we use a $4\times 4$ unit cell with three MZIs that is non-blocking over alternate pairs of output ports. One pair is indicated by the solid red arrows; the other pair is indicated by the dotted pink arrows. \textbf{\textsf{(c)}}~We connect unit cells like a binary tree, reducing four input lines to two output lines at each level. To maintain the non-blocking property over the output lines, we connect appropriate pairs of output lines at each level.}
\label{suppfig:MZI}
\end{figure}

Mach-Zehnder interferometer (MZI) arrays are versatile design tools that have been demonstrated in optical neural networks~\cite[][]{Shen_2017}, programmable photonic circuits~\cite[][]{Harris_2018,Bogaerts_2020} and other physical applications. MZI arrays are well-suited for implementing all-to-all connectivity in quantum routers, because MZIs can act as quantum switches between pairs of waveguides (Supplementary Fig.~\ref{suppfig:MZI}a). Figure~5 provides an example of how a quantum router can be built with an MZI array. The photon loss due to an MZI array is dependent on the array's depth, which is the maximum number of MZIs a photon has to traverse when travelling from one end of the array to the other. We want to design routers using shallow MZI arrays so as to minimize on-chip losses.

An $m$-register, $k$-way router will have $(k+2)$ output ports: $k$ of the ports will lead to each of the router's neighboring nodes, whereas the remaining two ports will go to detectors for local Bell state measurements. Combining an on-chip MZI with the two local detectors will give the detector station required to generate entanglement locally using the Barrett-Kok protocol because MZIs can also function as beamsplitters. We assume that $k$ remains fixed as $m$ varies, because the number of qubits per repeater node is likely to increase significantly with further developments in integrated photonics \cite[][]{Wan_2020}.

The MZI array for an $m$-register, $k$-way router must be able to connect any register to any output port. Furthermore, we must be able to simultaneously connect any pair of registers to the two detector ports, because the incoming photons must arrive at the beamsplitter at the same time in the Barrett-Kok protocol. More stringent forms of the latter `non-blocking' requirement have been extensively studied in classical switching networks~\cite[][]{Clos_1953}. A rectangular `cross-point' arrangement of MZIs~\cite[][]{Lee_2019} is completely non-blocking over all output ports, but the depth of the MZI array scales linearly with the number of registers $m$. The photon losses of such a linear-depth array may be prohibitive in the limit of large $m$. On the other hand, a binary MZI tree will be able to connect any register to any output port, but it does not satisfy the non-blocking condition at the two detector ports.

`Dilated' MZI networks~\cite[][]{Lee_2019} are fully non-blocking, with a depth that scales logarithmically with $m$. However, dilated networks achieve the non-blocking property by incorporating significant routing redundancy between switches. As a result, the number of switches required scales quadratically with $m$, like the rectangular `cross-point' array. MZI arrays with more MZIs are harder to fabricate and control, even if the array depth and hence photon loss remains small.

We propose a tree-based MZI array design with logarithmic depth and a linear number of MZIs. Such a design is possible because we only require a fixed subset of output ports to be non-blocking, not all ports. 

We use the $4\times 4$ unit cell shown in Supplementary Fig.~\ref{suppfig:MZI}b to build the MZI array. By choosing the state of each MZI appropriately, any pair of inputs from the left can be simultaneously connected with alternate pairs of outputs on the right. Starting from the $m$ registers at the inputs of the MZI array, we build a tree of $4\times 4$ unit cells by connecting four consecutive lines using a unit cell, producing two output lines in the next level of the tree (Supplementary Fig.~\ref{suppfig:MZI}c). We repeat the process at each level until we are left with two non-blocking output lines. This tree forms the register routing layer in Fig.~5 of the main text.

We complete the connection between the registers and the $(k+2)$ output ports using an interposer, which switches the two non-blocking outputs (and other extra lines) from the register routing layer into the network routing layer or the local Bell state measurement layer. The network routing layer is simply a binary tree of MZIs that connects the interposer to the $k$ outward-facing ports. The local Bell state measurement layer contains one MZI that acts as a beamsplitter for the Barrett-Kok protocol.

We now count the depth of the MZI array. A photon from one of the $m$ registers must travel through $\log_2 (m/2)$ unit cells to reach the two non-blocking output lines of the register routing layer. As each unit cell has a depth of two MZIs, the register routing layer has a depth of $2\log_2(m/2)$. As shown in Fig.~5, we only need one MZI for the interposer. The network routing layer needs a depth-$(\log_2 k)$ MZI tree to link $k$ output ports and the interposer. The local Bell state measurement layer only needs one MZI, and it is located in parallel to the network routing layer. Therefore, the above construction produces an MZI array of depth $2\log_2(m/2) + 1 + \max(\log_2 k, 1) = 2\log_2(m) + \log_2(k) -1$, which is logarithmic in $m$. (If $m$ and/or $k$ are not powers of 2, we take the ceiling of $\log_2(m)$ and/or $\log_2(k)$ where appropriate.)

The number of MZIs needed for this construction scales linearly with $m$. The register routing layer requires $(m/2-1)$ unit cells, each of which has three MZIs. The interposer and local Bell state measurement layer need one MZI each, and the network routing layer needs $(k-1)$ MZIs. Hence, the total number of MZIs in the array is $3(m/2-1) + 1 + (k-1) + 1 = 3m/2 + k - 1 = \mathcal{O}(m)$.

\begin{table*}[ht!]
\centering
\caption{Timeline of a clock cycle, as shown in Fig.~3. Representative times are derived from Ref.~\cite{Humphreys_2018}.} \label{supptab:timeline_clockcycle}
\setlength\tabcolsep{3pt}
\begin{tabular}{c|>{\centering\arraybackslash}p{3.5cm}|>{\centering\arraybackslash}p{3.5cm}} \hline
    \multirow{2}{*}{Step} & \multicolumn{2}{c}{Time taken} \\ \cline{2-3}
     & Router & Standard repeater \\ \hline \hline
     Initialization and first Barrett-Kok pulse & \multicolumn{2}{c}{$6\, \mu \text{s}$} \\ \hline
     $X$ gate and second Barrett-Kok pulse & \multicolumn{2}{c}{$50\, \text{ns}$} \\ \hline
     Time until distant entanglement heralded & \multicolumn{2}{c}{distance $\div$ speed of light in fiber} \\ \hline
     Matching phase & $\approx 10 \, \mu \text{s}$ & N/A \\ \hline
     Bell state measurements & \multicolumn{2}{c}{$\approx 20 \, \mu \text{s}$} \\ \hline
\end{tabular}
\end{table*}

\begin{table*}[ht!]
\centering
\caption{Timeline of a full left-right entanglement, as shown in Fig.~3. Representative times are derived from Ref.~\cite{Humphreys_2018}.} \label{supptab:timeline_entanglement}
\setlength\tabcolsep{3pt}
\begin{tabular}{c|>{\centering\arraybackslash}p{7.0cm}} \hline
     Router step & Time taken \\ \hline \hline
     Time to distant entanglements &  $t_\text{distant} / p_\text{distant}$ \\ \hline
     Matching phase = idling time & $\approx 10 \, \mu \text{s}$ \\ \hline
     Bell state measurements & $\approx 20 \, \mu \text{s}$ \\ \hline
     \multicolumn{2}{c}{} \\ \hline
     Standard step & Time taken \\ \hline \hline
     Time to left distant entanglement & $t_\text{distant} / p_\text{distant}$ \\ \hline
     Time to right distant entanglement = idling time & $t_\text{distant} / p_\text{distant}$ \\ \hline
     Bell state measurements & $\approx 20 \, \mu \text{s}$ \\ \hline
\end{tabular}
\end{table*}

In summary, our construction satisfies the non-blocking requirement at the two detector ports using an $\mathcal{O}(\log m)$-depth, $\mathcal{O}(m)$-switch MZI array, making it more efficient than standard approaches for designing photonic switches.

\section{Proposed Timeline} \label{app:timeline}
For completeness and ease of comparison, we detail a timeline for the router and standard repeater in Supplementary Table~\ref{supptab:timeline_clockcycle}. The majority of the clock cycle is occupied by the round-trip entanglement time (i.e. the time between the second Barrett-Kok pulse and the heralding of distant entanglement). The entanglement matching phase of the router clock cycle is expected to be small compared to the rest of the clock cycle, so the router and standard clock cycles are similar.

However, the idling time for a router is significantly shorter than that for a standard repeater. Supplementary Table~\ref{supptab:timeline_entanglement} details the timeline of a full left-right entanglement, from the first entanglement attempt to the Bell state measurement. The idling time, i.e. the time between the first and second distant entanglements, approaches zero in the router as the degree of multiplexing $m$ becomes large, whereas the idling time of the standard repeater remains constant at $t_\text{distant} / p_\text{distant}$. The longer idling time of the standard repeater causes more decoherence, and thus the left-right entanglement generated by the standard repeater has a lower fidelity than that of the router in the limit of high multiplexing.

\bibliography{references_limited}